\documentclass{emulateapj}
\shorttitle{EVLA Follow-up of a Nearby Galaxy from ALFA ZOA}
\shortauthors{T. McIntyre et al.}
\begin{document}

\title{Discovery and Follow-up of a Nearby Galaxy from the Arecibo Zone of Avoidance Survey}

\author{T. P. McIntyre,\altaffilmark{1,2} R. F. Minchin,\altaffilmark{2} E. Momjian,\altaffilmark{3} P. A. Henning,\altaffilmark{1} A. Kaur,\altaffilmark{4} B. Parton\altaffilmark{4}}
\altaffiltext{1}{Department of Physics and Astronomy, University of New Mexico, MSC07 4220, 1 Univ of New Mexico, Albuquerque, NM 87131-0001.}
\altaffiltext{2}{Arecibo Observatory, HC03 Box 53995, Arecibo, PR 00612.}
\altaffiltext{3}{NRAO, Domenici Science Operations Center, PO Box O, 1003 Lopezille Rd, Socorro, NM 87801.}
\altaffiltext{4}{Department of Physics and Astronomy, Clemson University, 118 Kinard Laboratory, Clemson, S.C. 29634.}

\begin{abstract}
The Arecibo L-Band Feed Array Zone of Avoidance (ALFA ZOA) Survey has discovered a nearby galaxy, ALFA ZOA J1952+1428, at a heliocentric velocity of +279 $\mathrm{km \; s^{-1}}$. The galaxy was discovered at low Galactic latitude by 21-cm emission from neutral hydrogen (H\,{\sc{i}}). We have obtained follow-up observations with the EVLA and the 0.9-m SARA optical telescope. The H\,{\sc{i}} distribution overlaps an uncataloged, potential optical counterpart. The H\,{\sc{i}} linear size is 1.4 kpc at our adopted distance of D = 7 Mpc, but the distance estimate is uncertain as Hubble's law is unreliable at low recessional velocities. The optical counterpart has m$_B$ = 16.9 mag and $B$ - $R$ = 0.1 mag. These characteristics, including M$_{HI}$ = 10$^{7.0}$ M$_\odot$ and $L_{B}$ = $10^{7.5}$ L$_\odot$, if at 7 Mpc, indicate that this galaxy is a blue compact dwarf, but this remains uncertain until further follow-up observations are complete. Optical follow-up observations are ongoing and near infrared follow-up observations have been scheduled.
\end{abstract}


\section{Introduction}
The Arecibo L-Band Feed Array Zone of Avoidance (ALFA ZOA) Survey searches for 21-cm line emission from neutral hydrogen (H\,{\sc{i}}) in galaxies behind the disk of the Milky Way. The survey  uses the ALFA receiver on the 305-m Arecibo Radio Telescope\footnote{Arecibo Observatory is part of the National Astronomy and Ionosphere Center, which, when the observations were made, was operated by Cornell University under a cooperative agreement with the NSF.}. This region of the sky is termed the Zone of Avoidance by extragalactic astronomers because of its low galaxy detection rate. Extragalactic observations at visual wavelengths struggle with high extinction levels. Near and far infrared observations suffer confusion with Galactic stars, dust, and gas. 21-cm line observations are sensitive to late-type galaxies in general and are not affected by extinction. As a spectral line survey, we generally only have confusion with Galactic H\,{\sc{i}} within approximately $\pm$100 $\mathrm{km \; s^{-1}}$. The ALFA ZOA survey is sensitive to galaxies behind the Milky Way that go undetected at other wavelengths. It has been suggested by Loeb and Narayan (2008) that undiscovered mass behind the Milky Way may explain the discrepancy between the cosmic microwave background dipole and what is expected from the gravitational acceleration imparted on the Local Group by matter in the local universe (Erdogdu et al. 2006).

Two large area H\,{\sc{i}} ZOA surveys have preceded ALFA ZOA; the Dwingeloo Obscured Galaxies Survey and the HI Parkes Zone of Avoidance Survey (HIZOA). The Dwingeloo survey detected 43 galaxies in the northern hemisphere within $\pm 5^\circ$ of the Galactic plane. It was sensitive only to nearby, massive objects because of its relatively high noise level of 40 mJy beam$^{-1}$ (with velocity resolution of 4 km s$^{-1}$; Henning et al. 1998). More recently, HIZOA covered decl. = -90 to +25 at 6 mJy/beam rms (with velocity resolution of 27 km/s), and detected about 1000 galaxies (Donley et al. 2005; Henning et al. 2000, 2005, Shafi 2008.)
    
The ALFA ZOA survey is being conducted in two phases: a shallow and a deep phase. The shallow phase (rms = 5 mJy with velocity resolution of 10 km/s) covers 900 square degrees through the inner Galaxy ($30^\circ<l<75^\circ$, $|b|<10^\circ$) and is expected to detect 500 galaxies. Hundreds of galaxies have been detected so far, and data reduction and analysis are ongoing. This is complemented by a deep survey ($30^\circ<l<75^\circ$, $170^\circ<l<215^\circ$, $|b|<5^\circ$), 5 times more sensitive, in which we expect to detect thousands of galaxies (based on the HIMF of Davies et al., 2011) but for which observations are not yet complete.

This paper presents the discovery and the results from follow-up observations of a nearby galaxy, ALFA ZOA J1952+1428. Section 2 describes the discovery and follow-up with the Arecibo Radio Telescope. Section 3 describes follow-up observations with the Expanded Very Large Array\footnote{The National Radio Astronomy Observatory (NRAO) is a facility of the National Science Foundation operated under cooperative agreement by Associated Universities, Inc.} (EVLA). Section 4 describes ongoing optical follow-up with the 0.9-m Southeastern Association for Research in Astronomy\footnote{The SARA-North telescope at Kitt Peak National Observatory is owned and operated by the SARA consortium.} (SARA) telescope. Section 5 discusses the results from these observations.

\section{Arecibo Observations and Results}
ALFA ZOA J1952+1428 was initially detected with the shallow portion of the ALFA ZOA survey. Observations were taken with the Mock spectrometer covering 300 MHz bandwidth in two 170 MHz sub-bands of 8192 channels each, giving a Hanning smoothed velocity resolution of 10 $\mathrm{km \; s^{-1}}$ at z = 0. The survey uses a meridian nodding mode observation technique: the telescope slews up and down in zenith angle along the meridian for an effective 8 second integration time per beam giving rms = 5 mJy per beam. Observations were taken in 2008 and 2009. The angular resolution of the survey is 3.4$^{\prime}$. More details of the ALFA ZOA survey techniques are presented by Henning et al. (2010).

In order to confirm this detection, it was followed up with the L-band Wide receiver on the Arecibo telescope for 180 seconds of integration time using a total power on-off observation. Data were taken with the WAPP spectrometer with 4096 channels across a bandwidth of 25 MHz giving a velocity resolution of 1.3 km\,s$^{-1}$ and rms = 2.5 mJy.

The spectrum from the follow-up observation can be seen in Figure 1. The velocity width at 50\% peak flux is $w_{50} = 28$ $\pm$ 2 $\mathrm{km \; s^{-1}}$. The heliocentric velocity measured at the mid-point of the velocity width is $v_{hel} = 279 \pm 1$ $\mathrm{km \; s^{-1}}$. The integrated flux density is $F_{HI}$ = 0.94 $\pm$ 0.07 Jy $\mathrm{km \; s^{-1}}$. Errors were calculated as in Henning et al. (2010) following the methods of Koribalski et al. (2004). ALFA ZOA J1952+1428 has no cataloged counterparts within $7^\prime$ (two Arecibo half-power beamwidths) in the NASA Extragalactic Database (NED). 

\begin{figure}
\plotone{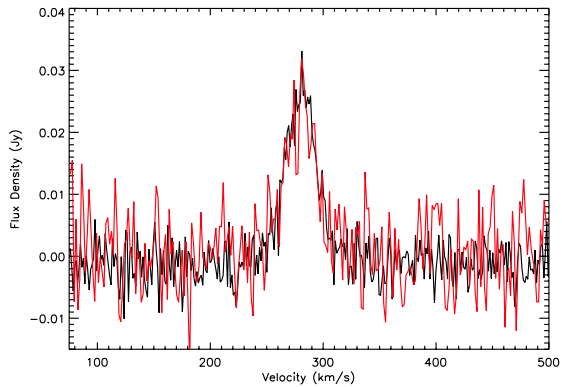}
\caption{Integrated H\,{\sc{i}} spectrum of ALFA ZOA J1952+1428. The black line is from a follow-up observation with the Arecibo telescope and the red line is from the EVLA observations. The EVLA recovered all H\,{\sc{i}} emission to within 1$\sigma$.}
\end{figure}

\section{EVLA Observations, Data Reduction, and Analysis}
Follow-up C-configuration EVLA observations were carried out to obtain high-resolution H\,{\sc{i}} imaging of ALFA ZOA J1952+1428. The observations were scheduled dynamically for 3 $\times$ 1 hour sessions and observed on December 3rd and 4th, 2010. We utilized the WIDAR correlator with 2 MHz bandwidth over 256 spectral channels, resulting in 7.8 kHz (1.6 $\mathrm{km \; s^{-1}}$) channel width. The on-source integration time was two hours. The source 3C48 was used to calibrate the flux density scale and the source J1925+2106, $9^\circ$ from the target source, was used to calibrate the complex gains. The editing, calibration, deconvolution, and processing of the data were carried out in AIPS. Line free channels were extracted from the spectral-line data cube and averaged to image the continuum in the field of the H\,{\sc{i}} source and to refine the phase and amplitude calibration. The resulting phase and amplitude solutions were applied to the spectral-line data set, and a continuum-free UV data cube was constructed by subtracting the continuum emission. We then created a total intensity (Stokes I) H\,{\sc{i}} image cube that was CLEANed using natural weighting giving a synthesized beamwidth of $15.13^{\prime\prime} \times 13.13^{\prime\prime}$ and an rms noise level of 2.6 mJy beam$^{-1}$  channel$^{-1}$. Moment 0 (H\,{\sc{i}} flux density), moment 1 (velocity field), and moment 2 (velocity dispersion) maps were produced from the H\,{\sc{i}} image cube by smoothing across 3 velocity channels (5 $\mathrm{km \; s^{-1}}$) and 5 pixels spatially ($20^{\prime\prime}$ at $4^{\prime\prime}$ per pixel) and clipping at 2.6 mJy (the 1$\sigma$ level of the unsmoothed cube). These maps can be seen in Figure 2.

The angular extent of the H\,{\sc{i}} out to 1 M$_\odot$ pc$^{-2}$ is $44^{\prime\prime} \times 40^{\prime\prime}$. The H\,{\sc{i}} flux density shows a main peak and a secondary peak $16^{\prime\prime}$ away that overlaps a region of high velocity as well as significant velocity dispersion. The velocity field shows structure but non-uniform rotation. The integrated flux from the Arecibo and the EVLA spectra are  0.94 $\pm$ 0.07 Jy km s$^{-1}$ and 0.80 $\pm$ 0.13 Jy km s$^{-1}$, respectively. The EVLA recovered all integrated flux to within 1$\sigma$. A comparison of the H\,{\sc{i}} profile between Arecibo and the EVLA can be seen in Figure 1.

\begin{figure}
\hspace{5.15mm}\resizebox{0.775\columnwidth}{!}{\includegraphics{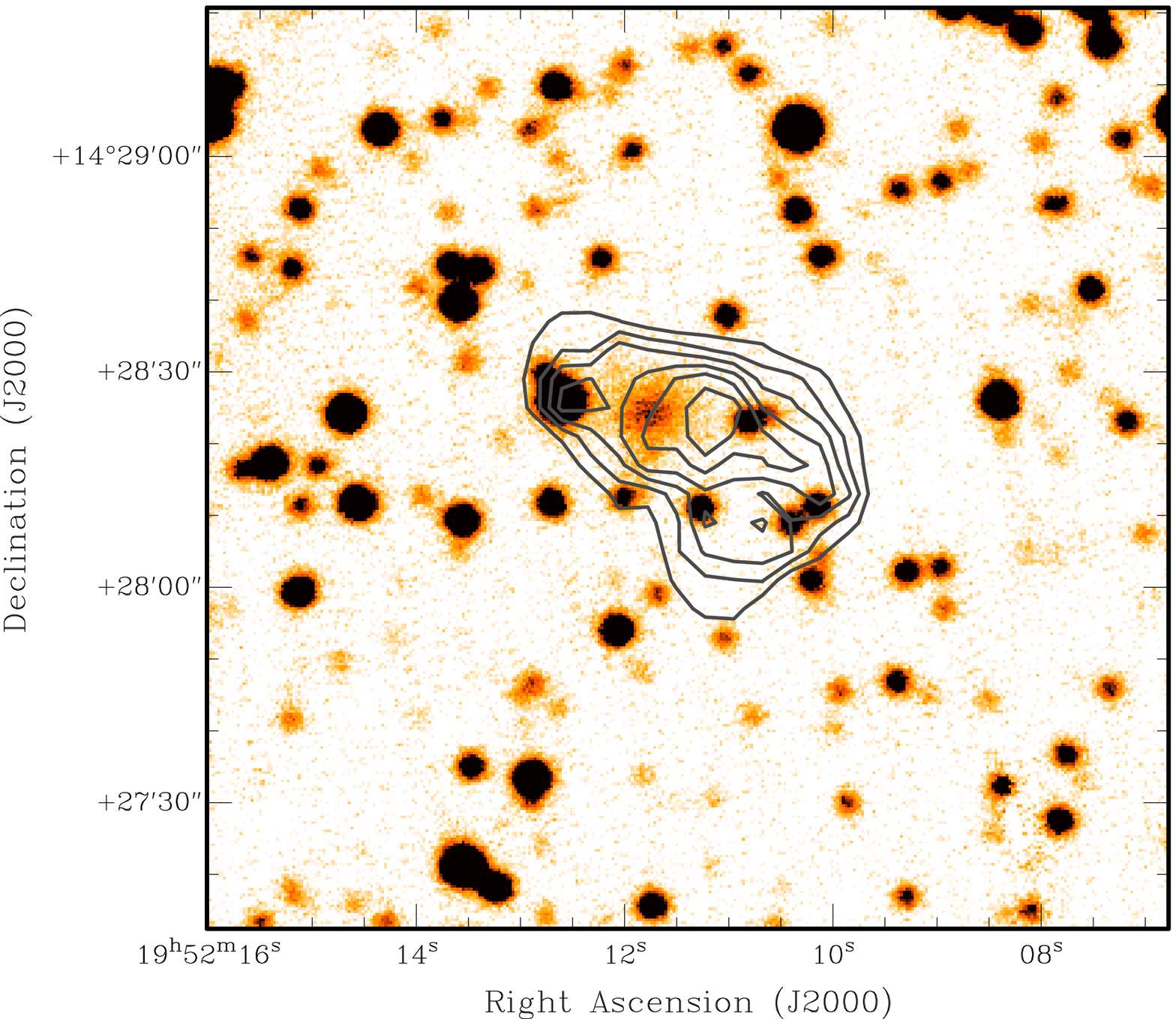}}\\
\resizebox{\columnwidth}{!}{\plotone{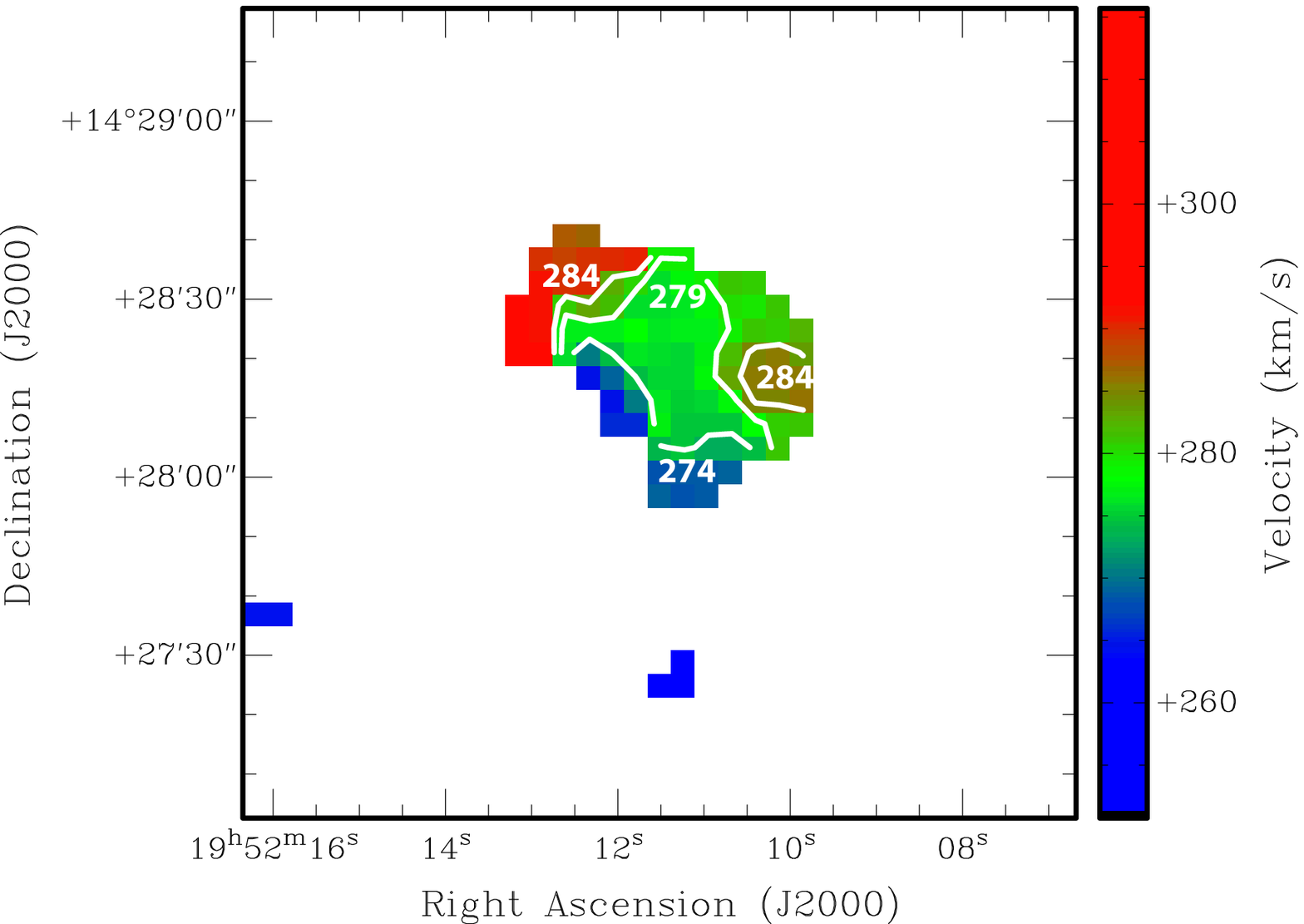}}\\
\hspace*{3.85mm}\resizebox{0.79\columnwidth}{!}{\plotone{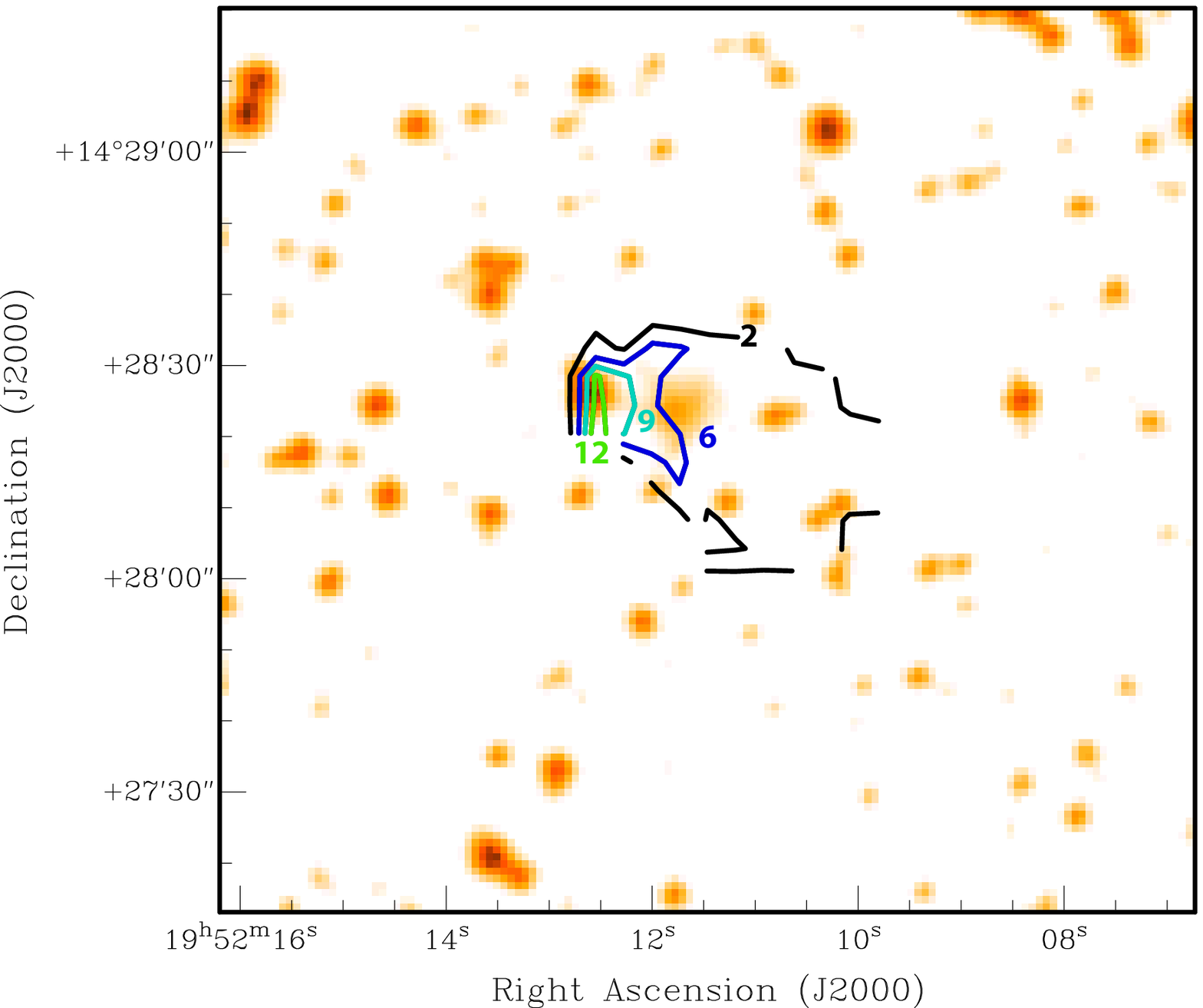}}
\caption{$2^\prime\times2^\prime$ SARA $B$ band image and EVLA moment maps. Top. H\,{\sc{i}} column density map overlayed on a SARA $B$ band image. Contours are at 1, 2, 3, 4, 5, 6 $\mathrm{\times 10^{20} cm^{-2}}$. The H\,{\sc{i}} peak is offset by $8.7^{\prime\prime}$ from the apparent optical counterpart. Center. Velocity field of ALFA ZOA J1952+1428 showing structure but not uniform rotation with contours at 274, 279, 284 $\mathrm{km \; s^{-1}}$. Bottom. Velocity dispersion map showing with contours at 2, 6, 9, 12 $\mathrm{km \; s^{-1}}$.}
\end{figure}

\section{Optical Observations, Data Reduction, and Analysis}
Digitized Sky Survey\footnote{The Digitized Sky Surveys were produced at the Space Telescope Science Institute under U.S. Government grant NAG W-2166. The images of these surveys are based on photographic data obtained using the Oschin Schmidt Telescope on Palomar Mountain and the UK Schmidt Telescope. The plates were processed into the present compressed digital form with the permission of these institutions.} (DSS) images show what looks to be a very faint, uncataloged galaxy that may be the optical counterpart. The DSS magnitudes of this object from SuperCOSMOS\footnote{This research has made use of data obtained from the SuperCOSMOS Science Archive, prepared and hosted by the Wide Field Astronomy Unit, Institute for Astronomy, University of Edinburgh, which is funded by the UK Science and Technology Facilities Council.} are m$_B$ = 17.5 $\pm$ 0.3 mag, m$_R$ = 17.0 $\pm$ 0.3 mag. The extinction in the area is relatively low for the ZOA with values estimated to be A$_B$ = 1.1 and A$_R$ = 0.7, from the DIRBE/$IRAS$ extinction maps (Schlegel, Finkbeiner, \& Davis 1998) though these values are somewhat uncertain at such low Galactic latitudes. Applying extinction corrections gives $B - R$ = 0.1 mag.

In order to obtain more accurate photometry, ALFA ZOA J1952+1428 was observed using a Bessell $B$ band filter on April 12, 2011 with the 0.9-m SARA telescope at Kitt Peak National Observatory using an Apogee Alta U 42 2048 $\times$ 2048 CCD. The field of view was 13.8$^\prime$ $\times$ 13.8$^\prime$, giving a plate scale of $0.4^{\prime\prime}$ pixel$^{-1}$. The source was low on the horizon with an average airmass of 1.7 and an average seeing of $2^{\prime\prime}$.  Nine 5-minute exposures were taken on source for a total exposure time of 45 minutes and calibration was done using the equatorial standard star PG1657+078A (Landolt 1992).

The CCD images were bias-subtracted, dark-corrected, flat-fielded and co-added in IRAF. The image can be seen in Figure 2. The APPHOT package was used for standard star photometry. The reduced image reached a 1$\sigma$ surface brightness level of 25 mag arcsec$^{-2}$. Astrometric calibration and aperture photometry of ALFA ZOA J1952+1428 were carried out interactively with the Graphical Astronomy and Image Analysis Tool (GAIA). Flux from the galaxy could be recovered out to a radius of $8^{\prime\prime}$, reaching a surface brightness of 23.5 mag arcsec$^{-2}$, after which stellar contamination became significant. The recovered flux within this radius was m$_B$ = $16.9 \pm 0.1$ magnitudes.

Optical follow-up observations with the SARA telescope are ongoing. $B$, $V$, and $R$ band and H-alpha observations are planned over the coming months. $B$, $V$, and $R$ band observations are being taken to a 1$\sigma$ surface brightness of 26.5 mag arcsec$^{-2}$. This should be sufficient to see low surface brightness features that correspond to the faint outer parts of a normal spiral and will allow us to measure the galaxy's diameter at the standard 25 mag arcsec$^{-2}$ level.

\section{Discussion}
\subsection{Distance}
ALFA ZOA J1952+1428 has a heliocentric velocity of $v_{hel}$ = +279 $\mathrm{km \; s^{-1}}$. Solving for its Local Group centered velocity using derivations of the solar motion with respect to the Local Group by Courteau $\&$ van den Bergh (1999) gives $v_{LG}$ = 491 $\mathrm{km \; s^{-1}}$. Using Hubble's Law with $H_{0}$ = 70 $\; \mathrm{km \; s^{-1} \; Mpc^{-1}}$ puts this source at a distance of 7 Mpc. However, Hubble's Law is not a reliable distance indicator here because the dispersion of peculiar velocities in the local universe ($v_{hel} < 10,000$ $\mathrm{km \; s^{-1}}$) is $\sigma = 298 \pm 34$ km s$^{-1}$ (Masters 2008). The galaxy is probably not closer than 3 Mpc as the H\,{\sc{i}} linear size at this distance would be smaller than most compact galaxies containing H\,{\sc{i}} (Huchtmeier et al. 2007). For the following analysis, we take the distance to be 7 Mpc, although future observations may well revise this number.

\subsection{Classification}
As can be seen in the EVLA and SARA images, the H\,{\sc{i}} peak is slightly offset ($\Delta\theta = 8.7^{\prime\prime}$) from the optical emission, indicating either a false counterpart or a disturbed H\,{\sc{i}} distribution. The offset is $\sim$300 pc at 7 Mpc, which is not uncommon even for isolated galaxies (c.f. $\sim$400 pc offset in VV 124, Bellazzini et al. 2011). This could conceivably be a pair of low-surface brightness dwarf galaxies (c.f. HIZSS 3 with separation of $\sim$900 pc, Begum et al. 2005), but there is no evidence for a second peak in the high signal-to-noise H\,{\sc{i}} spectrum shown in Figure 1. Further, ALFA ZOA J1952+1428 has half the velocity width that the pair in HIZSS 3 appeared to have; W$_{50}$ = 55 km s$^{-1}$ for HIZSS 3 (Henning et al. 2000) compared to W$_{50}$ = 28 km s$^{-1}$ here. Any second galaxy would have to be much closer both spatially and in velocity than the pair in HIZSS 3 in order to escape detection. Deeper interferometric observations would be needed to be entirely conclusive.

It is possible that ALFA ZOA J1952+1428 is a high velocity cloud (HVC) co-incident with an optical source (c.f. HIPASS J1328-30, Grossi et al. 2007), though this is unlikely as there is strong evidence that ALFA ZOA J1952+1428 is not an HVC. Its recessional velocity does not lie near HVCs in this part of the sky (c.f. Figure 3a in Morras et al. 2000). The nearest population of HVCs is the Smith Cloud which lies $10^{\circ}$ and 170 $\mathrm{km \; s^{-1}}$ away (Lockman et al. 2008) at its nearest point. If ALFA ZOA J1952+1428 were an HVC, it would be a remarkable outlier.  Furthermore, the velocity field of ALFA ZOA J1952+1428 shows a gradient ten times larger than those of HVCs (Begum et al. 2010).

ALFA ZOA J1952+1428 appears to be a dwarf galaxy judging from its Gaussian H\,{\sc{i}} profile and low H\,{\sc{i}} mass. At a distance of 7 Mpc, $M_{HI}$ = $10^{7.0}$ M$_\odot$, which is significantly lower than the gaseous content of spiral-type galaxies (Roberts and Haynes 1994). Also, its low luminosity ($L_{B}$ = $10^{7.5}$ L$_\odot$ at 7 Mpc), H\,{\sc{i}} content, and blue colors are strong evidence that it is not an early-type galaxy.

There is no possible counterpart visible in 2MASS archive images or listed within 8$^\prime$ in the 2MASS Extended Source Catalog (Jarrett et al. 2000). We plan follow-up NIR observations later this year with the 1.4-m Infrared Survey Facility in Sutherland, South Africa. We will use observations by its main instrument, SIRIUS, which has three detectors that operate simultaneously ($J$, $H$, $Ks$) with a field of view of 7.8 $\times$ 7.8 arcmin.

Table 1 summarizes the observational data and derived quantities. Columns 1 and 2 give equatorial coordinates (J2000) for the  H\,{\sc{i}} peak. Columns 3 and 4 give the Galactic coordinates. Column 5 gives the heliocentric velocity from the mid-point of the velocity width at 50\% peak flux. Column 6 gives the velocity width at 50\% peak flux. Column 7 gives the integrated flux. Column 8 gives the M$_{HI}$/L$_B$ ratio using the m$_B$ calculated from the SARA telescope. The error on L$_B$ is dominated by the unknown uncertainty in A$_B$, thus we do not quote an error on M$_{HI}$/L$_B$. Column 9 gives the angular size of the  H\,{\sc{i}} out to the 1 M$_\odot$ pc$^{-2}$ level. The last two Columns give values as a function of the distance to the galaxy in Mpc. Column 10 gives the linear size of the H\,{\sc{i}} at its largest extent. Column 11 gives total H\,{\sc{i}} mass.

\begin{table*}
\caption[]{Properties of ALFA ZOA J1952+1428.}
\label{tab:mba}
\scriptsize
\begin{tabular}{ccccccccccc}
R.A. & Decl. & $l$ & $b$ & $v_{hel}$ & $W_{50}$ &$F_{HI}$& M$_{HI}$/L$_B$ & H\,{\sc{i}} Size & $A_{o}$ & M$_{HI}$\\
(J2000.0) & (J2000.0) &  (deg) & (deg) & ($\mathrm{km \; s^{-1}}$) & ($\mathrm{km \; s^{-1}}$) &(Jy $\mathrm{km \; s^{-1}}$)& (M$_\odot$/L$_\odot$) & (arcsec) & (kpc) &(M$\odot$)\\
\tableline
19 52 11.8 & +14 28 24 &  52.82 & -6.42 & 279 $\pm$ 1  & 28 $\pm$ 2&0.94 $\pm$ 0.07& 0.3 & 44 $\times$ 40 & 0.21D &$10^{5.3}$D$^{2}$\\
\end{tabular}
\end{table*}

\subsection{Morphological Type}
Compared to other dwarf galaxies (Roberts \& Haynes 1994; O'Neil et al. 2000), ALFA ZOA J1952+1428 is not particularly gas rich, $M_{HI}/L_B$ = 0.3 M$_\odot$/L$_\odot$, but it is very small and blue with an H\,{\sc{i}} linear size of 1.4 $\times$ 1.3 kpc at 7 Mpc and $B$ - $R$ = 0.1 mag. The H\,{\sc{i}} mass, $\mathrm{M_{HI}/L_B}$ ratio, blue optical colors, and linear size of ALFA ZOA J1952+1428 are similar to those of blue compact dwarf (BCD) galaxies (Huchtmeier et al. 2007). BCDs are small, blue, irregular dwarf galaxies which have low surface brightness features, ongoing star formation, and higher metallicities than typical dwarf galaxies. The velocity field of ALFA ZOA J1952+1428 shows structure, but non-uniform rotation which is common in blue compact dwarf galaxies (Ramya et al. 2011). The velocity dispersion map shown in Figure 2 shows a significant amount of dispersion around the stellar looking object on the left side of the galaxy. This could be an ionized hydrogen region. H-alpha observations with the SARA telescope will examine this and quantify the star formation in this system. Deep $B$, $V$, and $R$ band observations with the SARA telescope will reveal whether there are low-surface brightness features in ALFA ZOA J1952+1428.

Alternatively, there is evidence for the existence of blue, metal-poor, gas-rich dwarf galaxies on the margins of galaxy groups (Grossi et al. 2007). These dwarfs are old (2 - 10 Gyrs) but have had remarkably little star formation in their history. They are thought to be galaxies in transition between dwarf irregular and dwarf spheroidal galaxies. ALFA ZOA J1952+1428 differs from the Grossi et al. galaxies because it has a lower $\mathrm{M_{HI}/L_B}$ ratio and appears to be a field galaxy, though it may be a part of a group behind the Milky Way that has not yet been discovered.

There is a recently discovered Local Group galaxy, VV124 (Bellazzini et al. 2011), which is similar to ALFA ZOA J1952+1428 in size, H\,{\sc{i}} mass, and $M_{HI}/L_B$ ratio. This galaxy is isolated, as ALFA ZOA J1952+1428 appears to be. There is only one galaxy (3.8$^\circ$ away) with $v_{hel} < 1000$ $\mathrm{km \; s^{-1}}$ within 10$^\circ$ of ALFA ZOA J1952+1428 in NED (though this is not unusual in the ZOA). VV124 also shows an offset between the H\,{\sc{i}} and the optical counterpart as well as a velocity field with structure but non-uniform rotation. VV124 is considered to be a precursor of modern dwarf spheroidal galaxies that did not undergo an interaction-driven evolutionary path. Follow-up observations will reveal whether ALFA ZOA J1952+1428 has metallicity and star formation rates similar to a VV124-type galaxy.

\subsection{Group Membership}
It is possible that ALFA ZOA J1952+1428 is a Local Group galaxy, but it appears unlikely; ALFA ZOA J1952+1428 does not follow the relationship between radial velocity and angle from the solar apex that most other Local Group members do, as can be seen in Figure 3 (Courteau $\&$ van den Bergh 1999). Further, the linear size of the H\,{\sc{i}} would be 210 pc at 1 Mpc, which is 4 times smaller than the smallest compact dwarfs (Huchtmeier et al. 2007).

\begin{figure}
\plotone{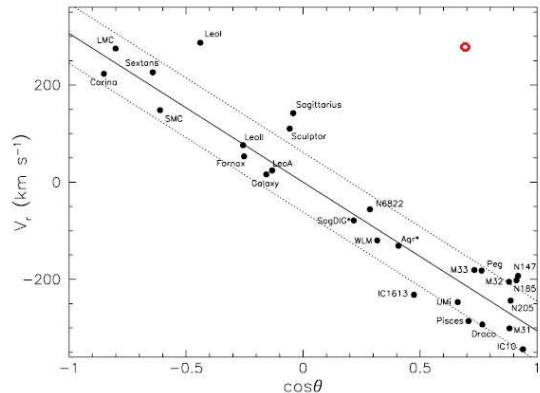}
\caption{Heliocentric velocity versus the cosine of the angular distance from the solar apex (modified with permission from Courteau and van den Bergh 1999). The black dots are Local Group galaxies and the red, open circle is ALFA ZOA J1952+1428. The solid line represents the solar motion solution of Courteau and van den Bergh ($v$ = 306 $\mathrm{km \; s^{-1}}$, $l = 99^\circ$, $b = -3^\circ$). The dotted lines represent the Local Group radial velocity dispersion, $\sigma_{r} = $ 61 $\mathrm{km \; s^{-1}}$.}
\end{figure}

There are no known galaxy groups within $60^{\circ}$ with $v_{hel} <$ 1000 $\mathrm{km \; s^{-1}}$ (Fouqu\'{e} et al. 1992), making ALFA ZOA J1952+1428 either a field galaxy or a member of an undiscovered nearby group. Continued analysis of the ALFA ZOA data may clarify this.

\section*{Acknowledgments}
\acknowledgments{TM gratefully acknowledges the NAIC Predoctoral Fellowship and the New Mexico Space Grant for supporting this work. The authors thank Dieter Hartmann for helping to obtain optical follow-up observations with the SARA telescope. The NASA/IPAC Extragalactic Database (NED) is operated by the Jet Propulsion Laboratory, California Institute of Technology, under contract with the National Aeronautics and Space Administration.}

\clearpage


\begin{thebibliography}{}

\bibitem[0(2000)]{Begum05} Begum, A., Chengalur, J. N., Karachentsev, I. D., \& Sharina, M. E. 2005, MNRAS, 359, 53

\bibitem[0(2000)]{Begum10} Begum, A., Stanimirovic, S., Peek, J.E., Ballering, N.P., Heiles, C., Douglas, K.A., Putman, M., Gibson, S.J., Grcevich, J., Korpela, E.J., Lee, M., Saul, D., \& Gallagher, J.S. 2010, ApJ, 722, 395

\bibitem[0(2000)]{Bellazzini11} Bellazzini, M., Beccari, G., Oosterloo, T. A., Galleti, S., Sollima, A., Correnti, M., Testa, V., Mayer, L., Cignoni, M., Fraternali, F., Gallozzi, S. 2011, A\&A, 527, 58

\bibitem[0(2000)]{Courteau99} Courteau, S., \& van den Bergh, S. 1999, AJ, 118, 337

\bibitem[0(2000)] {Davies11} Davies, J. I., et al. 2011, MNRAS, in press

\bibitem[0(2000)] {Donley05} Donley, J. L., et al. 2005, AJ, 129, 220

\bibitem[0(2000)]{Erdogdu06} Erdogdu, P., et al. 2006, MNRAS, 373, 45

\bibitem[0(2000)]{Fouque92} Fouqu\'{e}, P., Gourgoulhon, E., Chamaraux, P., \& Paturel, G. 1992, A\&AS, 93, 211

\bibitem[0(2000)]{Grossi07} Grossi, M., Disney, M. J., Pritzl, B. J., Knezek, P. M., Gallagher, J. S., Minchin, R. F., \& Freeman, K. C. 2007, MNRAS, 374, 107

\bibitem[0(2000)]{Haynes84} Haynes, M. P., Giovanelli, R., \& Chincarini, G. L. 1984, ARA\&A, 22, 445

\bibitem[0(2000)]{Henning98} Henning, P. A., Kraan-Korteweg, R. C., Rivers, A. J., Loan, A. J., Lahav, O., \& Burton, W. B. 1998, AJ, 115, 584

\bibitem[0(2000)]{Henning00} Henning, P. A., Staveley-Smith, L., Ekers, R. D., Green, A. J., Haynes, R. F., Juraszek, S., Kesteven, M. J., Koribalski, B., Kraan-Korteweg, R. C., Price, R. M., Sadler, E. M., \& Schr\"oder, A. 2000, AJ, 119, 2686

\bibitem[0(2000)]{Henning05} Henning, P. A., Kraan-Korteweg, R. C., \& Staveley-Smith, L. 2005, ASP Conf. Ser., 329, 199

\bibitem[0(2000)]{Henning10} Henning, P. A., Springob, C. M., Minchin, R. F., Momjian, E., Catinella, B., McIntyre, T. P., Day, F., Muller, E., Koribalski, B., Rosenberg, J. L., Schneider, S., Staveley-Smith, L., \& van Driel, W. 2010, AJ, 139, 2130

\bibitem[0(2000)]{Huchtmeier07} Huchtmeier, W. K., Petrosian, A., Gopal-Krishna, \& Kunth, D. 2007, A\&A 462, 919

\bibitem[0(2000)]{Jarrett00} Jarrett, T. H., Chester, T., Cutri, R., Schneider, S., Skrutskie, M., \& Huchra, J. P. 2000, AJ, 119, 2498

\bibitem[0(2000)]{Koribalski04} Koribalski, B. S. et al. 2004, AJ, 128, 16

\bibitem[0(2000)]{Landolt92} Landolt, A. U. 1992, AJ, 104, 340

\bibitem[0(2000)]{Lockman08} Lockman, F. J., Benjamin, R. A., Heroux, A. J., \& Langston, G. I. 2008, ApJ, 679, 21

\bibitem[0(2000)]{Loeb08} Loeb, A., \& Narayan, R. 2008, MNRAS, 386, 2221

\bibitem[0(2000)] {O'Neil05} O'Neil, K., Bothun, G. D., \& Schombert, J. 2000, AJ, 119, 1360

\bibitem[0(2000)]{Masters08} Masters, K. 2008, ASP Conf. Ser., 395, 137

\bibitem[0(2000)]{Morras00} Morras, R., Bajaja, E., Arnal, E. M., \& Poppel, W. G. L. 2000, A\&AS, 142, 25

\bibitem[0(2000)]{Ramya11} Ramya, S., Kantharia, N. G., \& Prabhu, T. P. 2011, ApJ, 728, 124

\bibitem[0(2000)]{Roberts94} Roberts, M. \& Haynes, M. 1994, ARA\&A, 32, 115.

\bibitem[0(2000)]{Schlegel98} Schlegel, D. J., Finkbeiner, D. P., \& Davis, M. 1998, ApJ, 500, 525

\bibitem[0(2000)]{Shafi05} Shafi, N. 2008, MSc thesis, Univ. of Cape Town


\end{thebibliography}
\end{document}